\documentclass[aps,prl,groupedaddress]{revtex4}

\usepackage{amsfonts,amssymb}
\usepackage{amsthm}
\usepackage{tikz}
\usetikzlibrary{calc}
\usetikzlibrary{arrows}
\usepackage{tikz-3dplot}
\usepackage{color} 
\usepackage[fleqn]{amsmath}
\usepackage{amsthm}
\usepackage{amssymb}
\usepackage{amsfonts}
\usepackage{bbm}
\usepackage{epsfig}
\usepackage{times}
\usepackage{hyperref}
\usepackage{bm}
\usepackage{float} 
\usepackage{appendix} 
\usepackage{mathrsfs}
\usepackage{appendix}
\usepackage{amsmath}%
\usepackage{MnSymbol}%
\usepackage{wasysym}%

\usepackage{graphicx}

\DeclareGraphicsExtensions{.png}

\newcommand{\comment}[1]{}

\theoremstyle{plain}
\newtheorem{theorem}{Theorem}
\newtheorem{lemma}{Lemma}

\theoremstyle{definition}

\begin{document} 
	
\title{Unconstrained Summoning for relativistic quantum information processing}

\author{Adrian \surname{Kent}} \affiliation{Centre for
	Quantum Information and Foundations, DAMTP, Centre for Mathematical
	Sciences, University of Cambridge, Wilberforce Road, Cambridge, CB3
	0WA, U.K.}  \affiliation{Perimeter Institute for Theoretical
	Physics, 31 Caroline Street North, Waterloo, ON N2L 2Y5, Canada.}
	\date{\today}

\begin{abstract} 
We define a summoning task to require propagating an 
unknown quantum state to a point in space-time belonging 
to a set determined by classical inputs at points in
space-time.  We consider the classical analogue, in
which a known classical state must be returned at
precisely one allowed point.  We show that, when the
inputs are unconstrained, any
summoning task that is possible in the classical
case is also possible in the quantum case. 
\end{abstract}

\maketitle
\section{Introduction}

In the not necessarily distant future, many or most significant
economic decisions may depend on algorithms that process
information arriving around the world as efficiently and
quickly as possible, given light speed signalling constraints.
Such algorithms will need to decide not only whether to make a
trade, but also when and where.   Quantum money enables 
efficient transactions, by allowing a delocalized quantum money token
effectively to be ``summoned'' to a given space-time point
as a result of information distributed through space-time. 
There are also alternative technological solutions
\cite{kent2018summonable} in the form of token schemes that do not require long term quantum state
storage and in some embodiments are based entirely on classical
communication.   

This paper defines types of summoning task that naturally arise
in relativistic economies, and shows that for an important and
natural class of these tasks the two technologies have the
same functionality.    That is, if quantum money can be 
summoned to the correct space-time point in a given scenario,
then so can classical tokens, and conversely.   
This has implications for the foundations of 
relativistic quantum theory, since summoning has proved
a very fruitful way of understanding the properties of 
quantum information in space-time
\cite{kent2013no,kent2012quantum,hayden2016summoning,adlam2016quantum}
and we also discuss these.    

\section{Summoning}

Summoning \cite{kent2013no, kent2012quantum} is a task defined between two parties, Alice and Bob,
who each have networks of collaborating agents. 
Each party trusts their own agents but not those of the other party.
Summoning tasks discussed in the literature to date take the following
form.  In some fixed background space-time, Bob creates a quantum state (say, a 
qudit) in pre-agreed physical form, keeping its classical description
private among his agents.   One of his agents gives the state to one
of Alice's at some point $P$.   At one of a number of pre-agreed later points 
$\{ c_i \}_{ i \in I}$, Bob's local agent may ask for the state back
from Alice's local agent.     The pre-agreed set $I$ may be finite or
infinite; the original discussion of the task \cite{kent2013no}
allowed
$I$ to contain every point in the future of $P$ and also considered
the possibility that $I$ contains just two space-like separated
points. 

Alice must then return the state at some point or within some region
related to $c_i$ in a pre-agreed way.   For example, in one version
considered in Ref. \cite{kent2012quantum}, a request requires the
state to be returned at the same point in space at a slightly later
time, with respect to a given frame. 
In Ref. \cite{hayden2016summoning}, this was generalized so that a request
at $c_i$ requires the state to be returned at some pre-agreed point
$r_i \succ c_i$.   We say such a summoning task is {\it possible} in
relativistic quantum theory if there is
an algorithm that guarantees that Alice will comply with any allowed
request (assuming she has ideal devices and instantaneous computing
operations).  Alice is allowed to return only a 
single state (here a qudit), so that compliance means
guaranteeing that this is the state originally received.   

Another interesting version of summoning allows
any non-zero number of valid return points \cite{adlam2016quantum}.  
In this case we say the task is possible if there is an algorithm that 
guarantees that Alice will return the state to some valid 
return point.  
For example, one can consider a variant of Hayden-May's version
of summoning in which requests may be made at any non-zero
number of call points $\{c_i \}_{i \in I'}$, where $I' \subseteq I$,
and the state 
must be returned at one of the corresponding return points,
i.e. at some $r_i$ where $i \in I'$. 
Interestingly, this is a strictly harder version of the task
\cite{adlam2016quantum}.

\section{Extending the definition of summoning}

For definiteness, we focus on the case of Minkowski space-time, 
which is a good local approximation in the region of the solar system.  
All the variants of summoning considered to date \cite{kent2013no,
  kent2012quantum,hayden2016summoning,adlam2016quantum}
are examples of a general class of quantum tasks in Minkowski
space-time \cite{kent2012quantum}, in which Alice receives some number
of classical and/or quantum inputs at a set of points in space-time
and must produce classical and/or quantum outputs at another (possibly
identical or overlapping) set of points, where the output points
and information depend in some prescribed way on the input points
and information.   
For the summoning tasks considered in
Refs. \cite{hayden2016summoning,adlam2016quantum}, 
the classical information consists of single bits ($1$ or $0$
corresponding
respectively to ``call'' or ``no
call'') at a finite set of ``call points'' $\{ c_i \}_{i=1}^n$, 
with a corresponding set of ``return points'' $\{ r_i \}_{i=1}^n$.
An input $1$ at $c_i$ gives the instruction that returning the 
state at $r_i$ is required \cite{hayden2016summoning} or is one
of the required set of possibilities \cite{adlam2016quantum}. 
The summoning tasks considered in Refs.  \cite{kent2013no,
  kent2012quantum} also have single bit inputs. 
However, more general types of input and more general rules 
for defining allowed return points are also interesting.  
The key feature common to all summoning tasks considered to date is 
that an unknown quantum state is supplied and must be returned
at some later point belonging to a set
that is identified by information supplied 
during the task. 
It is natural to extend the term summoning to cover all such tasks.

We focus here on the case in which all the information supplied,
other than the original unknown state, is classical, the sets of input
points and possible return points are both finite and known in advance, and finite bounds on the
information to be supplied at each input point are also known 
in advance.   Our terminology here is to be understood modulo these
restrictions, so that we speak here of ``summoning'' rather
than ``finite-input-point bounded-classical-input finite-output-point
summoning''.   This is purely for brevity: more general types of summoning are also
interesting.   The bounds on classical inputs are imposed to 
simplify the notation; our results extend to general classical inputs,
including unbounded integer or real number inputs. 

We thus define a {\it summoning task with one return
  point} in a given space-time to be a task set by one party, Bob, for
another party, Alice, who has an arbitrarily dense network of
collaborating agents. 
Bob gives Alice (i.e. Bob's local agent gives Alice's local agent)
an unknown quantum state at the start point $P$.   
Alice must produce the state at some return point
$Q \succ P$.   Here $Q \in \{ Q_1 , \ldots , Q_N \}$.
Its identity depends, according to
pre-agreed rules, on 
classical information supplied at some set of input points $\{ P_i
\}_{i=1}^M$.   The point $P$ and the sets $\{ P_i \}$ and $\{ Q_i \}$ are known to both
parties in advance of the
task.  The sets may overlap, and may include $P$.
We take the sets of input and return points to be finite, although 
the infinite case is also interesting. 
The classical information Bob supplies at $P_i$ is an 
integer $m_i$ in the range $0 \leq m_i \leq ( n_i - 1 )$.  
Alice must return the state to a point $Q(
m_1, \ldots , m_M )  \in \{ Q_1 , \ldots , Q_N \}$.
Alice knows the functional dependence of $Q$ and the 
set  $\{ n_i \}_{i=1}^M$ in
advance of the task; she learns the values of the $m_i$ only
at the points $P_i$.   
To exclude trivial cases that complicate the notation, we
assume that every return point may be designated by some
set of inputs: i.e. for each $i$ there is at least one set of allowed
inputs $m_1, \ldots , m_M $ such that 
$Q_i = Q (m_1, \ldots , m_M )$.
We also assume that every input takes more than one possible value,
i.e. that  $n_i \geq 2$ for each $i$. 
We say the task is {\it possible} if, 
given unlimited predistributed classical and/or quantum resources, 
ideal devices and instantaneous classical and quantum computational
power, there is an algorithm by which Alice can guarantee 
to return the state to $Q( m_1 , \ldots , m_M )$ for any 
allowed set of inputs $\{ m_1 , \ldots , m_M \}$.   

We define a {\it summoning task with at most one return point}
similarly, allowing that $Q(m_1 , \ldots , m_M )$ may be any point 
in the pre-agreed set $\{ Q_i \}_{i=1}^N$ or may be the empty set $\emptyset$, in 
which case the state should not be returned at any point. 
Such tasks are possible only if there is an algorithm by which
Alice can guarantee that
the state is returned to a point $Q_i$ if and only if 
$Q_i = Q( m_1 , \ldots , m_M )$, and is not returned anywhere if 
$ Q( m_1 , \ldots , m_M ) = \emptyset$. 

We may also define a {\it summoning task with multiple
return points}, for which $Q(m_1 , \ldots , m_M )$ may be
any subset of $\{ Q_i \}_{i=1}^M$, and $| Q(m_1 , \ldots , m_M )| >
1$ for at least one set of inputs.\footnote{A fully exhaustive
terminology would again distinguish between the cases where 
$Q(m_1 , \ldots , m_M )$ is empty for at least one set of inputs 
and where it is never empty.  We will not need to discuss these
separately here.}
Here we 
assume that for each $i$ there is at least one set of allowed
inputs $m_1, \ldots , m_M $ such that 
$Q_i \in Q (m_1, \ldots , m_M )$.
Such tasks are possible only if the algorithm guarantees that
the state is returned to a point $Q_i \in Q(m_1 , \ldots , m_M )$ if and only if 
$Q( m_1 , \ldots , m_M )$ is non-empty, and is not returned anywhere if 
$ Q( m_1 , \ldots , m_M ) = \emptyset$.

\begin{figure}[h]
\centering
\includegraphics[scale=0.4]{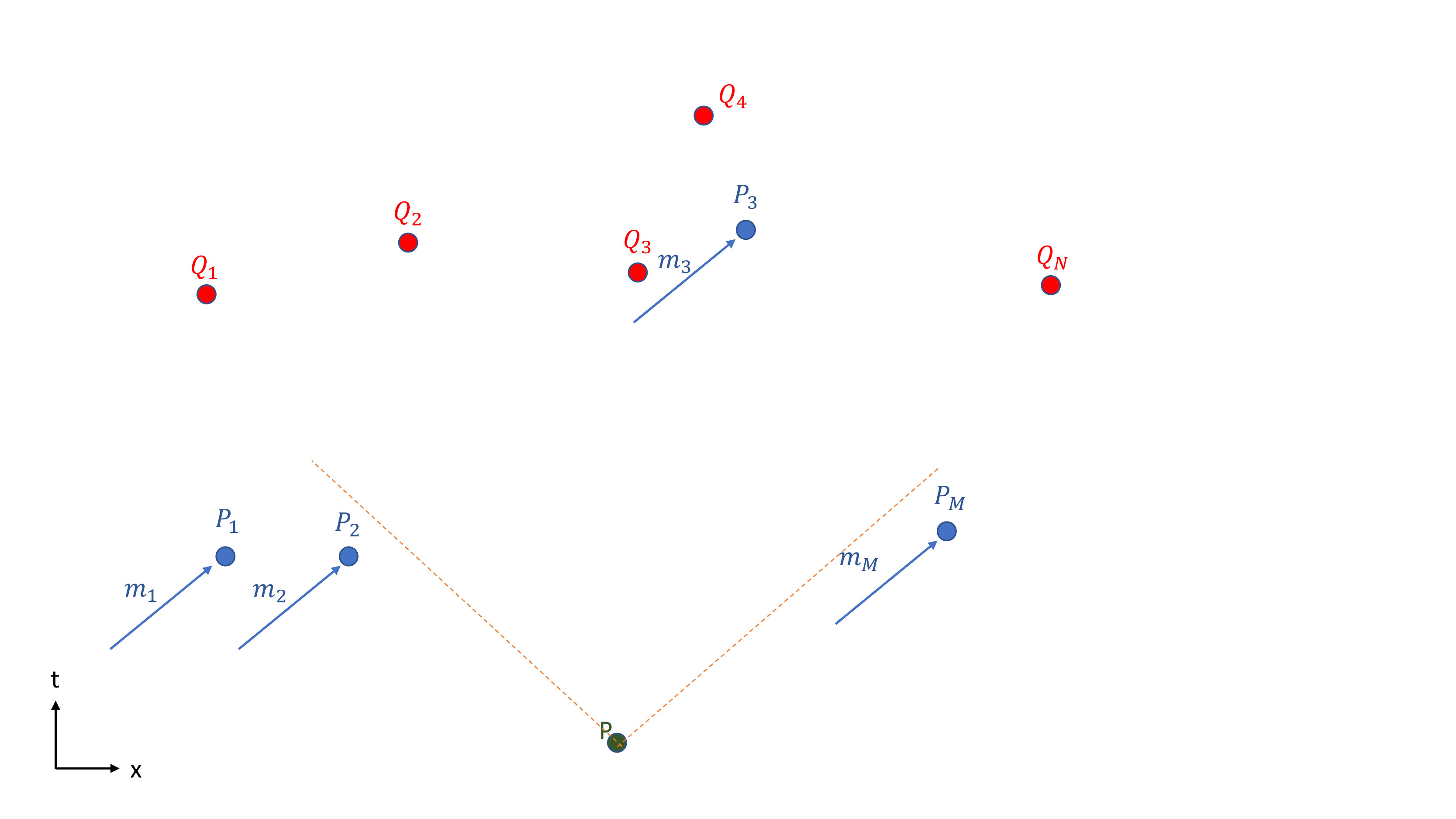}
\caption{A summoning task with at most one return point.  
Alice is given an unknown quantum state at the point $P$ 
and classical data in the form of integers $m_i$ at
the points $P_i$.   The dashed lines indicate the
future light cone of $P$.  
She is required to send the quantum state 
to $Q(m_1 , \ldots , m_M ) \in \{ Q_1 \ldots Q_M \}$, 
or to retain it if $Q(m_1 , \ldots , m_M ) = \emptyset$.
}\end{figure}

We distinguish between summoning tasks with
{\it constrained inputs} and those with {\it unconstrained inputs}.
In the former, Alice is guaranteed that some non-trivial constraint holds   
on the possible inputs. 
That is, there is at least one set of inputs
 $\{ m_1 , \ldots , m_M \}$, in the prescribed
ranges $0 \leq m_i \leq (n_i -1 )$, that is guaranteed never
to arise.   Hayden-May \cite{hayden2016summoning} define a 
summoning task with constrained inputs, since only one $1$
input is allowed.  The version of multi-call summoning
\cite{adlam2016quantum} in which it is also allowed that no call
may be made at any call point is a task with unconstrained inputs.

\section{Classical summoning tasks}

It is also interesting to consider classical versions of these
summoning tasks.   Here Alice is given a classical state  at point
$P$.   We assume she can perfectly determine its classical description
instantaneously,  broadcast the description everywhere at light
speed, and reconstruct a perfect copy instantaneously anywhere she
receives the broadcast description.  She thus may make arbitrarily
many perfect copies of the state at $P$ or anywhere within its 
causal future.  At each point $Q_j (j \in J)$ she must declare if she
is returning the state; if she does, she must supply a copy of it. 
We assume she only has classical resources.  
We say the task is {\it classically possible} if there is an
algorithm which guarantees that she will return a copy of the
state at precisely one valid return point (if there are any)
and at no other point.  (So, if there are no valid return points,
she does not return the state anywhere.) 

Not all classically possible summoning tasks are also possible
when a quantum state is being summoned.  
For example, in the original setting for the no-summoning theorem \cite{kent2013no},
Alice knows that precisely one call will be made, but does not 
know where.   If she is required to return a classical state, she 
may simply broadcast its description everywhere, and return it in 
response to the call, wherever it is made.   
The same is true of Hayden-May summoning \cite{hayden2016summoning}.
So long as all the return points are in the causal future of 
the start point, and the call points are in the causal past of
each return point, Alice knows at each return point whether or not the
state should be returned.  Again, she may simply broadcast the 
description of the classical state to all the return points, and
return it in response to the call, regardless of the relationship
between the causal diamonds defined by the call and return points.
In both cases, the task is thus classically possible, but quantum
summoning is not generally possible.  

These examples show that classical summoning may be possible
while quantum summoning is impossible for a summoning task with constrained inputs. 
We now focus on summoning tasks with unconstrained inputs.

\section{Summoning tasks with unconstrained inputs and at most one
  return point}

Consider now a classically possible 
summoning task with unconstrained inputs and at most one return point.

\lemma Each return point $Q_i$ is in the causal future of the start
point $P$, i.e. $Q_i \succeq P$ for each $i$. 

\proof Each return point $Q_i$ may  be designated by some set of
allowed inputs.   If $Q_i$ is so designated, Alice must propagate
the qudit from $P$ to $Q_i$.   Unless $Q_i \succeq P$, this would require superluminal
signalling. $\blacksquare$

\lemma  For every pair of return points $( Q_i , Q_j )$ the set of 
common past input points $S_{ij} = \{ P_k : P_k \preceq Q_i \, \&  \, P_k \preceq Q_j
\}$ is non-empty.

\proof Both $Q_i$ and $Q_j$ are designated return points for some 
(different) sets of inputs.  Since the task is classically possible, 
the inputs at points in $S_i = \{ P_k : P_k \preceq Q_i \}$ must determine
whether or not $Q_i$ is a (and hence the) valid return point.  
Since both $Q_i$ and $Q_j$ are valid return points for some sets
of inputs, and the task is classically possible, $S_{ij}$ cannot
be empty, otherwise some sets of inputs on $S_i$ and $S_j$ would
be consistent with both $Q_i$ and $Q_j$ being valid return points. $\blacksquare$

\lemma  For every pair of return points $Q_i , Q_j$, any
possible set of inputs at their common past input points $S_{ij}$ 
must logically exclude at least one of the pair as the designated
return point.  

\proof If there is a set of inputs
at points in $S_{ij}$ that is consistent with both $Q_i$ and $Q_j$ being a valid
return point, these inputs must form part of a complete set of inputs
that is consistent with both $Q_i$ and $Q_j$ being valid 
return points, since the inputs are unconstrained and it must
be knowable at each $Q_k$ whether $Q_k$ is a valid return
point.  This contradicts the assumption of at most
one valid return point.  $\blacksquare$

The next result relies on the technique of distributed non-local
computation \cite{vaidman2003instantaneous}, which relies on
using pre-shared entanglement to implement a series of 
teleportation operations, with the classical teleportation
data broadcast so that it is available to reconstruct the
required quantum state at the appropriate point. 

In the simplest example, Alice is given a quantum state $\psi$ 
at point $P$ and inputs $m_1 , m_2$ at points $P_1 , P_2$ 
respectively, and is required to return $\psi$ at point
$Q (m_1 , m_2 ) \in \{ Q_1 , Q_2 \}$, where the function
$Q( m_1 , m_2 )$ is known in advance.   
Suppose that $Q_i \succeq P$ and $Q_i \succeq P_j$ for $i=1,2$
and $j=1,2$.  She may then carry out a teleportation
operation on  $\psi$ at $P$ using a predistributed entangled
state shared between $P$ and (say) $P_1$, broadcasting the
classical teleportation data.   This produces a 
quantum state $\rho_1$ at $P_1$ from which $\psi$ may later be
reconstructed.   Now suppose that Alice has pre-shared
$n_1$ labelled entangled states between $P_1$ and $P_2$.
At $P_1$, she carries out a teleportation
operation on $\rho_1$, using the entangled state with
label $m_1$, and broadcasts the classical teleportation data
and the value of $m_1$.   At $P_2$, she sends the entangled state
with label $m$ to the point $Q(m , m_2 )$, and broadcasts
the value of $m_2$. 
At $Q_1$ and $Q_2$, she receives both $m_1$ and $m_2$, and so 
knows which is the required return point.   She also receives
all the relevant teleportation data and the required quantum
state at that return point, and so can reconstruct and
return $\psi$ there.

By iterating this technique, we obtain the next lemma.
This requires a large amount of pre-shared entanglement
in examples with many call points; we are concerned
here only with feasibility in principle, rather than
resource optimization.

\lemma If a state $\psi_{ij}$ is initially located at
the start point $P$, it may be propagated in such a way that it
arrives at $Q_i$ if the inputs on $S_{ij}$ preclude $Q_j$ 
(but not $Q_i$) as a valid return point and at $Q_j$ if the inputs on $S_{ij}$ preclude
$Q_i$ (but not $Q_j$).   

\proof  A non-local computation \cite{vaidman2003instantaneous}, taking into account the inputs
at all points $P_k \in S_{ij}$, can be carried out by iterative
teleportations from $P$ to one of these points and through a complete
sequence of the remaining points, 
such that the teleportation output containing the quantum
information linked with $\psi_{ij}$ is propagated appropriately
to $Q_i$ or $Q_j$.   We have that $P \preceq Q_i$ and $P_k \preceq
Q_i$ for all $P_k \in S_{ij}$, and similarly for $Q_j$. The state may thus be reconstructed at 
the relevant return point from the classical teleportation data. $\blacksquare$

This algorithm could be extended to give rules as to where
the state goes if both $Q_i$ and $Q_j$ are precluded; however this is
irrelevant for our purposes.

The next lemma uses techniques of and results about quantum secret sharing, developed 
in Refs. \cite{gottesman2000theory,smith2000quantum}.
A {\it quantum secret sharing scheme} for 
quantum states in a given Hilbert space $H$ is defined by a
quantum operation $A:H \rightarrow H_1 \otimes \ldots \otimes H_n$ from $H$ to a 
tensor product of component Hilbert spaces $H_i$, together with 
quantum operations $A_S : \otimes_{i \in S} H_i \rightarrow H$ for
some subsets $S \subseteq \{ 1 , \ldots , n \}$, such that 
$A_S \cdot A = I$.   The {\it access structure} of the secret
sharing scheme is the list $L$ of subsets $S$ for which 
such an operation $A_S$ is defined.   Effectively, a quantum
secret sharing scheme allows an unknown quantum state $\psi$
to be shared among $n$ parties in such a way that any subset
$S$ of them belonging to the access structure may reconstruct
$\psi$.   Quantum secret sharing schemes can be constructed
with any access structure that is monotonic (so that if $S_1 \in L$
and
$S_1 \subseteq S_2$ then $S_2 \in L$) and  does not violate the no-cloning
theorem  (so that if $S_1 \cap S_2 = \emptyset$ and $S_1 \in L$ then
$S_2 \notin L$).   Here again we are presently interested only in whether
a scheme exists in principle; we do not consider resource
optimization. 

\lemma  There is a quantum secret sharing scheme for a general
state $\psi$, with components
labelled by $\{ \psi_{ij} \}_{1 \leq i < j \leq N}$, with the property
that $\psi$ may be reconstructed from any subset of the form
$\{ \psi_{ij} \}_{1 \leq j \leq N ; j \neq i }$, where we set
$\psi_{ji} = \psi_{ij}$ if $j>i$. 

\proof No pair of these subsets are disjoint, and so they
satisfy the conditions to generate an access structure for a quantum
secret sharing scheme \cite{gottesman2000theory,smith2000quantum}. $\blacksquare$

\theorem A classically possible 
summoning task with unconstrained inputs and at most one return point
is also a possible quantum summoning task.

\proof This follows by construction from the preceding lemmas.   $\blacksquare$

\section{Summoning tasks with unconstrained inputs and multiple return points}

Consider now a classically possible 
summoning task with unconstrained inputs and multiple return
points.   Since the task is classically possible, there must be 
some classical algorithm that allows Alice to decide, based on
the inputs, to exclude
returning at one of any pair $Q_i$ and $Q_j$ of return points
when both are valid.   Without loss of generality, we may
assume this algorithm is deterministic, since Alice has only
classical resources, and any classical randomness in 
a probabilistic algorithm may be precomputed and predistributed. 
The algorithm must be consistent: i.e. it must identify a 
valid return point when there is one.  The determination of whether $Q_i$ or $Q_j$
is excluded can only depend on the inputs at points in $S_{ij}$, by
causality.  

We can thus incorporate the algorithm within a refined definition of the task, 
producing a summoning task with unconstrained inputs 
and at most one return point.   We may delete any return points that
are never used by the algorithm.   This defines a possible quantum
summoning task, by the previous discussion. 
Hence we have:

\begin{theorem} A classically possible 
summoning task with unconstrained inputs and multiple return points
is also a possible quantum summoning task.
\end{theorem}

\section{Classical and Quantum Possibility}

We have shown that classically possible summoning tasks with
unconstrained inputs are 
also possible quantum summoning tasks, 
whether they have at most one return point or multiple return points. 

Conversely, consider a quantum summoning task that can be solved with a 
deterministic quantum algorithm, by which we mean an algorithm that
always returns the state to the same return point $Q(m_1 , \ldots, m_M
)$ for a given set of inputs $m_1 , \ldots , m_M$. 
By including appropriate ancillae, we can describe any such quantum
algorithm as a deterministic sequence of unitary operations, in
which unitaries $U_Q$ act on the collective state at point $Q$ and
propagate outputs along secure channels to further points $R_i \succ
Q$.  
All the operations that might
be performed in this algorithm can be 
described classically: 
``apply unitary $U_Q$ at $Q$'', ``prepare state $\phi$ at $Q'$'', and
so on, where each operation depends in a prescribed way on inputs
received at the relevant point or in its causal past.  
If the algorithm is deterministic in the sense above,
then the success of a summoning task that is supposed to 
propagate a state from the start point $P$ to a 
valid return point $Q_j$ is determined by, and 
deducible from, the subset of these operations applied 
in the past light cone of $Q_j$.  

Hence any deterministic quantum algorithm that guarantees
success may be simulated by classical communications,
describing the relevant operations and state
preparations, broadcast from the corresponding
space-time points.   This broadcast simulation allows 
Alice's agent at the valid return point $Q_j$ 
(if there is a valid return point), to deduce that the quantum
algorithm would have returned the quantum state there,
and thus that she should return a copy of the classical
state there.  It also allows Alice's agents at all other
return points $Q_k \neq Q_j$ to deduce that the quantum
algorithm would not have returned the quantum state
at their locations, and hence that they should not return copies
of the classical states.   This gives us:

\begin{theorem} A quantum summoning task with unconstrained
inputs and multiple return points that can be solved with a 
deterministic quantum algorithm is also classically possible. 
\end{theorem}

\section{Discussion}

Summoning can be thought of as a type of distributed quantum
computation, or more generally a sub-routine within such a
computation, in which quantum states need to be propagated
in response to incoming classical information.  
This information could come from nature (for example 
detected photon fluxes near given spacetime points), 
from human activity (for example local market prices 
at given points in time on a distributed financial
network), or as outputs from other computations or 
other parts of the same computation (which 
may themselves use natural and/or human-made inputs).

Our results further illustrate
(cf. \cite{vaidman2003instantaneous,kent2011unconditionally,kent2012quantum,
kent2011quantum,buhrman2014position,hayden2016summoning,adlam2016quantum})
the power of quantum
information in a relativistic context.
Roughly speaking, they show that if a classical observer at a given spacetime
point can know, from available classical data that has no known
constraints, that a given unknown quantum state should have been propagated to  
them, then there was an infallible algorithm that 
could have done so, and vice versa.  We take this as further support for viewing
summoning as a key primitive of relativistic quantum information
theory. 

The algorithm defined is almost certainly far from optimal for
most interesting unconstrained summoning tasks.   It would 
be very interesting to understand better how to optimize the 
use of entanglement and other resources for these tasks.
One strong motivation for doing so comes from financial
and other applications of distributed algorithms on networks
where relativistic signalling constraints are significant.
As noted earlier, token or money schemes that prevent 
illegitimate duplication can in principle be based on
quantum money, but can also be implemented by alternative 
techniques that require no long term quantum state storage,
and in some cases no quantum information processing at all \cite{kent2018summonable}.
Unconstrained summoning tasks are natural problems in this context:
a market agent wants to be able to respond as flexibly and fast
as possible to incoming market data across the network, and to 
be able to present their money token at the optimal point in
space-time, according to some appropriate financial metric.
An unconstrained quantum summoning scheme in which quantum
money tokens are summoned and presented is an elegant
theoretical solution.   However if, as seems very plausible,
such schemes generally require unfeasible amounts of entanglement
to solve realistic problems, then rival technologies
\cite{kent2018summonable} are likely to prove advantageous. 
\vskip 10pt

{\bf Acknowledgments} \qquad This work was partially supported by 
UK Quantum Communications Hub grant no. EP/M013472/1 and by 
Perimeter Institute for Theoretical Physics. Research at Perimeter
Institute is supported by the Government of Canada through Industry
Canada and by the Province of Ontario through the Ministry of Research
and Innovation.   
	
\bibliographystyle{unsrtnat}
\bibliography{gensummoning}{}

\end{document}